\documentclass[12pt]{article}
\usepackage{graphicx}

\def\pbnr{}
\def\speaker{Liu Zhiqiang}
\def\onbehalfof{BESIII Collaboration}
\def\title{Study of charmonium decays at BESIII}
\def\affiliation{Institute of High Energy Physics\\
Chinese Academy of Sciences, Beijing, China, 100049}
\def\support{The workshop was supported by the University of Manchester, IPPP, STFC, and IOP}

\newcommand\pubnumber{\pbnr}
\newcommand\pubdate{\today}
\def\Title#1{\begin{center} {\Large #1 } \end{center}}
\def\Author#1{\begin{center}{ \sc #1} \end{center}}

\newcommand{\OnBehalf}[1]{\sbox0{#1}\ifdim\wd0=0pt
        {}
	\else
	{\\on behalf of #1}
	\fi}
\newcommand{\SupportedBy}[1]{\sbox0{#1}\ifdim\wd0=0pt
        {}
	\else
	{\footnote{#1}}
	\fi}
\def\Address#1{\begin{center}{ \it #1} \end{center}}

\newcommand\pubblock{\includegraphics[width=5cm]{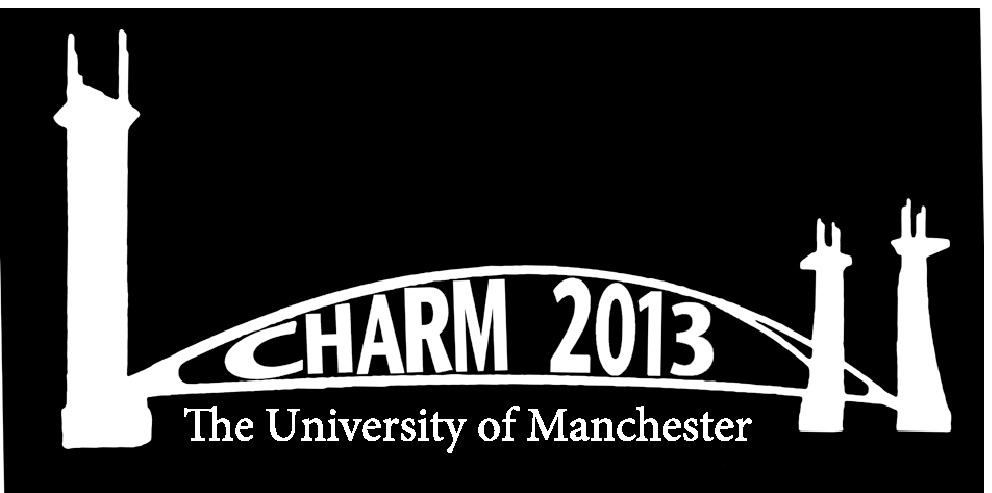}\hfill{\begin{tabular}{l} \pubnumber\\
         \pubdate  \end{tabular}}}
\newenvironment{Abstract}{\begin{quotation}  }{\end{quotation}}
\newenvironment{Presented}{\begin{quotation} \begin{center} 
             PRESENTED AT\end{center}\bigskip 
      \begin{center}\begin{large}}{\end{large}\end{center} \end{quotation}}
\def\Acknowledgements{\bigskip  \bigskip \begin{center} \begin{large}
             \bf ACKNOWLEDGEMENTS \end{large}\end{center}}
\def\venue{The 6$^{th}$ International Workshop on Charm Physics\\
(CHARM 2013)\\
Manchester, UK,  31 August -- 4 September, 2013}

\def\beq{\begin{equation}}
\def\eeq#1{\label{#1}\end{equation}}
\def\eeqn{\end{equation}}

\def\beqa{\begin{eqnarray}}
\def\eeqa#1{\label{#1}\end{eqnarray}}
\def\eeqan{\end{eqnarray}}

\let\bar=\overbar

\def\Dslash{\not{\hbox{\kern-4pt $D$}}}
\def\dslash{\not{\hbox{\kern-2pt $\del$}}}

\def\msb{{\bar{\ssstyle M \kern -1pt S}}}

\begin{document}
\begin{titlepage}
\pubblock

\vfill
\Title{\title}
\vfill
\Author{\speaker\SupportedBy{\support}\OnBehalf{\onbehalfof}}
\Address{\affiliation}
\vfill
\begin{Abstract}
Based on the data samples collected at the energies of $J/\psi$, $\psi(3686)$,
$\psi(3770)$ and $\psi(4040)$ resonances with the BESIII detector at the BEPCII
storage ring, we present charmonium decays in two aspects. One is the search for
baryonic decays of $\psi(3770)$ and $\psi(4040)$, including
$\Lambda \bar\Lambda\pi^+\pi^-$, $\Lambda\bar\Lambda\pi^0$,
$\Lambda \bar\Lambda\eta$, $\Sigma^+ \bar\Sigma^-$, $\Sigma^0 \bar\Sigma^0$,
$\Xi^-\bar\Xi^+$ and $\Xi^0\bar\Xi^0$. None are observed, and upper limits
are set at the 90\% confidence level. The other is the light hadron spectroscopy
from charmonium radiative decays, including the spin-parity analysis of the
$p\bar p$ mass-threshold enhancement in $J/\psi$ radiative decays and the
$\eta\eta$ system in $J/\psi\to\gamma\eta\eta$ radiative decays.
\end{Abstract}
\vfill
\begin{Presented}
\venue
\end{Presented}
\vfill
\end{titlepage}
\def\thefootnote{\fnsymbol{footnote}}
\setcounter{footnote}{0}
%

\section{Introduction}
The world largest datasets in $\tau$-charm energy region have been 
accumulated with the BESIII dectector at the BEPCII collider.
Based on the data samples of 225 million $J/\psi$ decays, 106 million
$\psi(3686)$ decays, $2.9$ fb$^{-1}$ at the $\psi(3770)$ resonance
and $482$ pb$^{-1}$ at the $\psi(4040)$ resonance, we present
two aspects of charmonium decays at BESIII. One is the search for
baryonic decays of $\psi(3770)$ and $\psi(4040)$,
including $\Lambda \bar\Lambda\pi^+\pi^-$, $\Lambda\bar\Lambda\pi^0$,
$\Lambda \bar\Lambda\eta$, $\Sigma^+ \bar\Sigma^-$,
$\Sigma^0 \bar\Sigma^0$, $\Xi^-\bar\Xi^+$ and $\Xi^0\bar\Xi^0$.
The other is the light hadron spectroscopy from charmonium radiative
decays, including the partial wave analysis results of the $p\bar p$
mass-threshold enhancement and $\eta\eta$ system in $J/\psi$ radiative
decays.

\section{Search for baryonic decays of $\psi(3770)$ and $\psi(4040)$}
The $\psi(3770)$ and $\psi(4040)$, broad $c\bar c$ resonances above
$D\bar D$ threshold, decay quite abundantly into open-charm final
states. While charmless decays of the $\psi(3770)$ and $\psi(4040)$
are possible, their branching fractions are supposed to be highly
suppressed.

The BES Collaboration measured the branching fraction
for $\psi(3770)$ decay to non-$D\bar D$ to be $(15\pm5)\%$ by using
different methods \cite{plb659_74,prd76_122002,prl97_121801,plb641_145}
under the hypothesis that only one simple $\psi(3770)$ resonance exists
in the center-of-mass energy region from 3.70 to 3.87 GeV.
The CLEO Collaboration obtained the branching fraction $\mathcal
B(\psi(3770)\to$ non-$D\bar D)=(-3.3\pm1.4^{+6.6}_{-4.8})$\%
\cite{prl104_159901} under the assumption that the interference of the
resonance decay, $\psi(3686)\to\gamma^*\to q\bar q\to hadrons$, with
the continuum annihilation, $\gamma^*\to q\bar q\to hadrons$, is
destructive at $\sqrt s=3.671$ GeV and constructive at $\sqrt s=3.773$ GeV
\cite{prl96_092002}. 

The BES Collaboration observed the first non-$D\bar D$ decay,
$\psi(3770)\to\pi^{+}\pi^{-}J/\psi$, with a branching fraction of
$(0.34\pm0.14\pm0.09)\%$ \cite{plb605_63}. The CLEO Collaboration
confirmed the same hadronic transition~\cite{prl96_082004}, and
observed other hadronic transitions $\pi^{0}\pi^{0}J/\psi$, $\eta
J/\psi$ \cite{prl96_082004}, and radiative transitions
$\gamma\chi_{cJ}(J=0,1)$ \cite{prl96_182002,prd74_031106} to
lower-lying charmonium states, and the decay to light hadrons
$\phi\eta$ \cite{prd73_012002}. While BES and CLEO have continued
to search for exclusive non-$D\bar D$ decays of $\psi(3770)$,
the total non-$D\bar D$ exclusive components are less than 2\%
\cite{pdg2012}. Meanwhile, there are fewer experimental measurements
for $\psi(4040)$ charmless decays. The BESIII Collaboration observed
for the first time the production of $e^+e^-\to\eta J/\psi$ at
$\sqrt s=4.009$ GeV. Assuming the $\eta J/\psi$ signal is from
a hadronic transition of the $\psi(4040)$, the fractional transition
rate is determined to be $\mathcal B(\psi(4040)\to\eta J/\psi)=
(5.2\pm0.5\pm0.2\pm0.5)\times10^{-3}$ \cite{prd86_071101}.
Search for other exclusive non-$D\bar D$ decays of $\psi(3770)$
and $\psi(4040)$ is urgently needed.

\begin{figure}[htb]
\centering
\includegraphics[width=3.5in,height=3.5in]{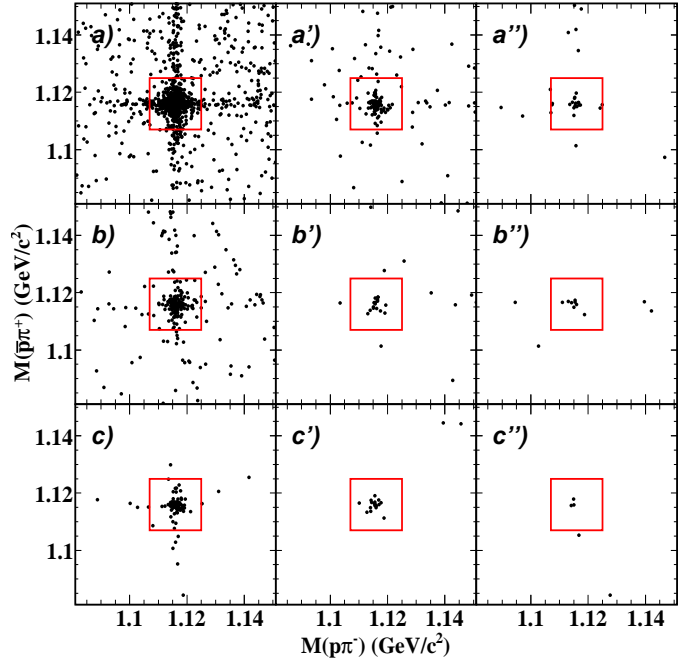}
\caption{Invariant mass of $p\pi^-$ versus $\bar p\pi^+$ distributions for
  $\Lambda\bar\Lambda\pi^+\pi^-$ [(a), (a') and (a'')], $\Lambda\bar\Lambda\pi^0$
  [(b), (b') and (b'')], $\Lambda\bar\Lambda\eta$ [(c), (c') and (c'')].
  The rectanglar regions indicate signal regions. The figures on the left
  (middle, right) side: data at $\sqrt s=3.773$ [4.009, continuum (3.543,
  3.554, 3.561, 3.600 and 3.650)] GeV.}
\label{fig:2D_1}
\end{figure}

\begin{figure}[htb]
\centering
\includegraphics[width=3.5in,height=4.5in]{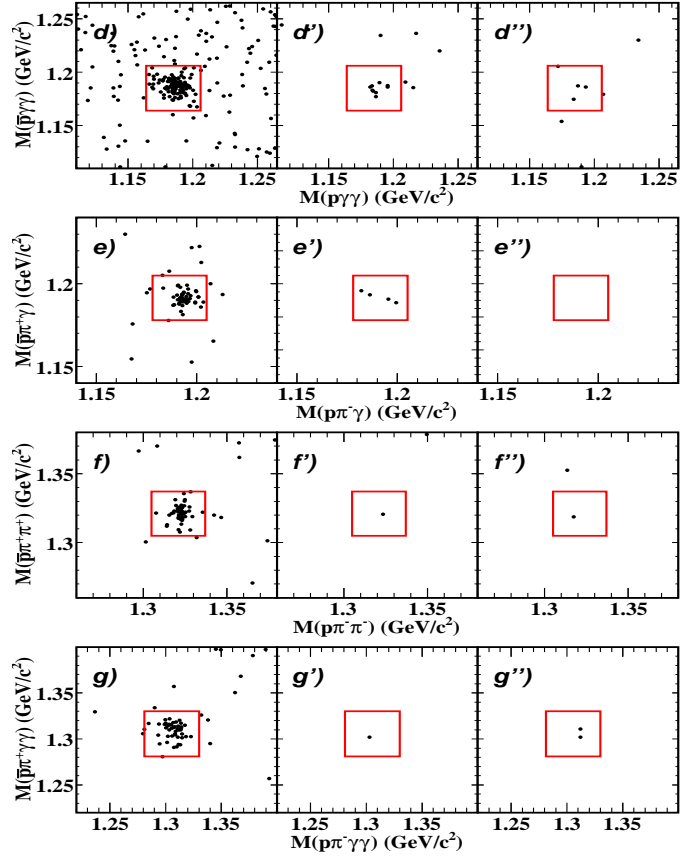}
\caption{Invariant mass of $p\gamma\gamma$, $p\pi^-\gamma$, $p\pi^-\pi^-$
  or $p\pi^-\gamma\gamma$ versus $\bar p\gamma\gamma$, $\bar p\pi^+ \gamma$,
  $\bar p\pi^+\pi^+$ or $\bar p\pi^+\gamma\gamma$ distributions for
  $\Sigma^+\bar\Sigma^-$ [(d), (d') and (d'')], $\Sigma^0\bar\Sigma^0$
  [(e), (e') and (e'')], $\Xi^-\bar\Xi^+$ [(f), (f') and (f'')],
  $\Xi^0\bar\Xi^0$ [(g), (g') and (g'')].}
\label{fig:2D_2}
\end{figure}

By analyzing data samples of $2.9$ fb$^{-1}$ collected at $\sqrt
s=3.773$ GeV, $482$ pb$^{-1}$ collected at $\sqrt s=4.009$ GeV and
$67$ pb$^{-1}$ collected at $\sqrt s=3.542$, 3.554, 3.561, 3.600 and
3.650 GeV, the BESIII experiment report the results of searches for
baryonic decays of $\psi(3770)$ and $\psi(4040)$, including final
states with the baryon pairs ($\Sigma^+ \bar\Sigma^-$,
$\Sigma^0 \bar\Sigma^0$, $\Xi^-\bar\Xi^+$, $\Xi^0\bar\Xi^0$) and
other $\Lambda\bar\Lambda$X modes (X = $\pi^+\pi^-$, $\pi^0$ and
$\eta$) \cite{prd87_112011}. In Figs.~\ref{fig:2D_1}
and \ref{fig:2D_2}, the two dimensional scatter plots are shown for each mode.

We assume that there is no interference between continuum production and
the $\psi(3770)$/$\psi(4040)$ resonance decay to the same baryonic
final state. We give the branching fractions $\mathcal{B}_{\psi(3770)/\psi(4040)\to f}$
and the upper limits $\mathcal{B}^{up}$ of $\psi(3770)$/$\psi(4040)$
baryonic decays for each mode in Table~\ref{tab:sig_psipp}. Since the
available continuum data is limited, the dominant error on each of the
seven branching fractions is from the continuum subtraction.

\begin{table}[t]
\begin{center}
\caption{ For each mode $f$, the following quantities are given:
the number of observed events $N_{obs}^f$ and background events $N_{B}^f$
at 3.773, 4.009 and 3.650 GeV; the scale factors $f_{co}^{3.773}$ and
$f_{co}^{4.009}$; the branching fractions ${\mathcal B}_{\psi(3770)\to f}$
and ${\mathcal B}_{\psi(4040)\to f}$, and the branching fraction upper limits
${\mathcal B}^{up}$ at 90\% C.L.}
\renewcommand{\arraystretch}{1.3} \footnotesize
\begin{tabular}{l|ccccccc}  
\hline \hline
&$N_{obs}^f$&$N_{B}^f$&$N_{obs}^f$&$N_{B}^f$& &${\mathcal B}_{\psi(3770)\to f}$
&${\mathcal B}^{up}$ \\
Mode $f$&(3.773)&(3.773)&(3.650)&(3.650)&$f_{co}^{3.773}$&[$\times 10^{-4}$]
&[$\times 10^{-4}$]  \\ \hline 
$\Lambda\bar\Lambda\pi^+\pi^-$  &$844.0\pm33.6$&$5.2$&$14.2^{+5.6}_{-4.2}$&$0.1$&$45.27$&$ 1.80^{+1.74}_{-2.30}\pm0.40$&$<4.7$ \\
$\Lambda\bar\Lambda\pi^0$       &$124.9\pm14.4$&$3.4$&$7.1^{+5.0}_{-2.2} $&$0.0$&$42.50$&$-1.28^{+0.67}_{-1.51}\pm0.15$&$<0.7$ \\
$\Lambda\bar\Lambda\eta$        &$74.0 \pm9.5 $&$0.9$&$3.0^{+3.6}_{-1.6} $&$0.0$&$44.76$&$-1.22^{+1.44}_{-3.21}\pm0.19$&$<1.9$ \\
$\Sigma^+\bar\Sigma^-$          &$100.5\pm11.9$&$0.7$&$3.3^{+4.3}_{-1.7} $&$0.1$&$38.27$&$-0.21^{+0.63}_{-1.56}\pm0.05$&$<1.0$ \\
$\Sigma^0\bar\Sigma^0$          &$43.5 \pm6.7 $&$0.0$&$0.0^{+2.2}_{-0.0} $&$0.0$&$38.69$&$ 0.30^{+0.05}_{-0.58}\pm0.05$&$<0.4$ \\
$\Xi^-\bar\Xi^+$                &$48.5 \pm7.0 $&$0.0$&$0.5^{+2.8}_{-1.4} $&$0.0$&$41.74$&$ 0.31^{+0.66}_{-1.32}\pm0.05$&$<1.5$ \\
$\Xi^0\bar\Xi^0$                &$43.5 \pm6.6 $&$1.3$&$2.0^{+3.2}_{-1.2} $&$0.0$&$40.13$&$-0.80^{+1.03}_{-2.72}\pm0.14$&$<1.4$ \\
\hline \hline
&$N_{obs}^f$&$N_{B}^f$&$N_{obs}^f$&$N_{B}^f$& &${\mathcal B}_{\psi(4040)\to f}$
&${\mathcal B}^{up}$ \\
Mode $f$&(4.009)&(4.009)&(3.650)&(3.650)&$f_{co}^{4.009}$&[$\times 10^{-4}$]
&[$\times 10^{-4}$]  \\ \hline 
$\Lambda\bar\Lambda\pi^+\pi^-$  &$79.2\pm10.0       $&$20.0$&$14.2^{+5.6}_{-4.2}$&$0.1$&$7.69$&$-3.57^{+2.45}_{-3.21}\pm0.79$&$<2.9$ \\
$\Lambda\bar\Lambda\pi^0$       &$14.5^{+4.1}_{-4.3}$&$0.5 $&$7.1^{+5.0}_{-2.2} $&$0.0$&$6.80$&$-2.14^{+0.97}_{-2.14}\pm0.28$&$<0.9$ \\
$\Lambda\bar\Lambda\eta$        &$16.0^{+4.2}_{-4.3}$&$3.6 $&$3.0^{+3.6}_{-1.6} $&$0.0$&$7.38$&$-1.60^{+2.06}_{-4.43}\pm0.57$&$<3.0$ \\
$\Sigma^+\bar\Sigma^-$          &$8.5^{+3.0}_{-3.2} $&$0.2 $&$3.3^{+4.3}_{-1.7} $&$0.1$&$4.92$&$-0.74^{+0.89}_{-2.14}\pm0.17$&$<1.3$ \\
$\Sigma^0\bar\Sigma^0$          &$4.0^{+3.2}_{-1.9} $&$0.0 $&$0.0^{+2.2}_{-0.0} $&$0.0$&$5.03$&$ 0.28^{+0.23}_{-0.79}\pm0.04$&$<0.7$ \\
$\Xi^-\bar\Xi^+$                &$1.0^{+2.2}_{-0.8} $&$0.0 $&$0.5^{+2.8}_{-1.4} $&$0.0$&$5.61$&$-0.21^{+0.94}_{-1.81}\pm0.04$&$<1.6$ \\
$\Xi^0\bar\Xi^0$                &$1.0^{+2.2}_{-0.8} $&$0.0 $&$2.0^{+3.2}_{-1.2} $&$0.0$&$5.36$&$-2.22^{+1.55}_{-3.93}\pm0.37$&$<1.8$ \\
\hline \hline
\end{tabular}
\label{tab:sig_psipp} 
\end{center}
\end{table}

\section{Light hadron spectroscopy in charmonium radiative decays}
\subsection{Partial wave analysis of $J/\psi\to\gamma p\bar p$}
An anomalously strong $p\bar p$ mass-threshold enhancement was first observed
by the BESII experiment in the radiative decay process $J/\psi\to\gamma p\bar p$
\cite{prl91_022001} and was recently confirmed by the BESIII \cite{cpc34_421}
and CLEO-c \cite{prd82_092002} experiments. The observation of the $p\bar p$  
mass-threshold enhancement also stimulated an experimental analysis of
$J/\psi\to\gamma\pi^+\pi^-\eta'$ decays, in which a $\pi^+\pi^-\eta'$ resonance,
the $X(1835)$, was first observed by the BESII experiment \cite{prl95_262001}
and recently confirmed with high statistical significance by the BESIII experiment 
\cite{prl106_072002}. Whether or not the $p\bar p$ mass-threshold enhancement
and $X(1835)$ are related to the same source still needs further study.
Spin-parity determinations and precise measurements of the masses, width, and
branching ratios are especially important. 

The BESIII experiment reports the first partial wave analysis (PWA) of the
$p\bar p$ mass-threshold structure produced via the decays of $J/\psi\to\gamma p\bar p$
and $\psi(3686)\to\gamma p\bar p$ \cite{prl108_112003}. Data samples containing
$(225.2\pm2.8)\times10^6$ $J/\psi$ events and $(106\pm4)\times10^6$ $\psi(3686)$
events accumulated with BESIII dector are used. The PWA of $J/\psi\to\gamma p\bar p$
and $\psi(3686)\to\gamma p\bar p$ are performed. In $J/\psi$ radiative decays,
the near-threshold enhancement $X(p\bar p)$ in the $p\bar p$ invariant mass is 
determined to be a $0^{-+}$ state. With the inclusion of Julich-FSI effects,
the mass, width and product of BRs for the $X(p\bar p)$ are measured to be:
$M=1832^{+19}_{-5}$(stat)$^{+18}_{-17}$(syst)$\pm19$(model) MeV/$c^2$,
$\Gamma=13\pm39$(stat)$^{+10}_{-13}$(syst)$\pm4$(model) MeV/$c^2$
(a total width of $\Gamma<76$ MeV/$c^2$ at the 90\% C.L.) and
BR[$J/\psi\to\gamma X(p\bar p)$]BR[$X(p\bar p)\to p\bar p$]$=$
[$9.0^{+0.4}_{-1.1}$(stat)$^{+1.5}_{-5.0}$(syst)$\pm2.3$(model)]
$\times10^{-5}$, respectively. 

\subsection{Partial wave analysis of $J/\psi\to\gamma\eta\eta$}
In accordance with the lattice QCD predictions \cite{prd73_014516,jhep_1210_170},
the lowest mass glueball with $J^{PC}=0^{++}$ is in the mass region from
1.5 to 1.7 GeV/$c^2$. However, the mixing of the pure glueball with nearby 
$q\bar q$ nonet mesons makes the identification of the glueballs difficult
in both experiment and theory. Radiative $J/\psi$ decay is a gluon-rich
process and has long been regarded as one of the most promising hunting
grounds for glueballs. In particular, for a $J/\psi$ radiative decay to 
two pseudoscalar mesons, it offers a very clean laboratory to search for
scalar and tensor glueballs because only intermediate states with $J^{PC}=even^{++}$
are possible. 

Based on the sample of $2.25\times10^8$ $J/\psi$ events collected with 
the BESIII detector, a full partial wave analysis on $J/\psi\to\gamma\eta\eta$
\cite{prd87_092009} was performed using the relativistic covariant tensor
amplitude method. The results show that the dominant $0^{++}$ and $2^{++}$
components are from the $f_0(1710)$, $f_0(2100)$, $f_0(1500)$, $f'_2(1525)$,
$f_2(1810)$, $f_2(2340)$. The resonance parameters and branching fractions
are also presented in Table~\ref{tab:etaeta}.

\begin{table}[t]
\begin{center}
\caption{ Summary of the PWA results, including the masses and widths for
resonances, branching ratios of $J/\psi\to\gamma X$, as well as the significance.
The first errors are statistical and the second ones are systematic. The
statistic significances here are obtained according to the changes of the log
likelihood. }
\renewcommand{\arraystretch}{1.3} \footnotesize
\begin{tabular}{l|cccc}  
\hline \hline
Resonance & Mass(MeV/$c^2$) & Width(MeV/$c^2$) &
${\mathcal B}(J/\psi\to\gamma X\to\gamma\eta\eta)$ & Significance \\ \hline 
$f_0(1500)$  & $1468^{+14+23}_{-15-74}$ & $136^{+41+28}_{-26-100}$ & $(1.65^{+0.26+0.51}_{-0.31-1.40})\times10^{-5}$ & 8.2 $\sigma$  \\
$f_0(1710)$  & $1795\pm6^{+14}_{-25}$   & $172\pm10^{+32}_{-16}$   & $(2.35^{+0.13+1.24}_{-0.11-0.74})\times10^{-4}$ & 25.0 $\sigma$ \\
$f_0(2100)$  & $2081\pm13^{+24}_{-36}$  & $273^{+27+70}_{-24-23}$  & $(1.13^{+0.09+0.64}_{-0.10-0.28})\times10^{-4}$ & 13.9 $\sigma$ \\
$f_2'(1525)$ & $1513\pm5^{+4}_{-10}$    & $75^{+12+16}_{-10-8}$    & $(3.42^{+0.43+1.37}_{-0.51-1.30})\times10^{-5}$ & 11.0 $\sigma$ \\
$f_2(1810)$  & $1822^{+29+66}_{-24-57}$ & $229^{+52+88}_{-42-155}$ & $(5.40^{+0.60+3.42}_{-0.67-2.35})\times10^{-5}$ & 6.4 $\sigma$  \\
$f_2(2340)$  & $2362^{+31+140}_{-30-63}$& $334^{+62+165}_{-54-100}$& $(5.60^{+0.62+2.37}_{-0.65-2.07})\times10^{-5}$ & 7.6 $\sigma$  \\
\hline \hline
\end{tabular}
\label{tab:etaeta} 
\end{center}
\end{table}

\section{Summary}
The BESIII experiment has collected the world's largest data samples at 
the $J/\psi$, $\psi(3686)$, $\psi(3770)$ and $\psi(4040)$ resonances.
Based on these samples, we present the charmonium decays at
BESIII. Search for baryonic decays of $\psi(3770)$ and $\psi(4040)$,
including $\Lambda \bar\Lambda\pi^+\pi^-$, $\Lambda\bar\Lambda\pi^0$,
$\Lambda \bar\Lambda\eta$, $\Sigma^+ \bar\Sigma^-$,
$\Sigma^0 \bar\Sigma^0$, $\Xi^-\bar\Xi^+$ and $\Xi^0\bar\Xi^0$ are very helpful
for understanding the non-$D\bar D$ decays of particles above $D\bar D$ threshold.
The partial wave analysis of the $p\bar p$ mass-threshold enhancement and 
$\eta\eta$ system in $J/\psi$ radiative decays are very meanful for understanding
the nature of the enhancement and finding the promising glueballs.

\Acknowledgements
I am grateful to the H91J11307A and Y21151025E funds in Institute of
High Energy Physics suporting me to attend the meeting.


\begin{thebibliography}{99}
\bibitem{plb659_74}
M. Ablikim {\it et al.} (BES Collaboration), Phys. Lett. B
{\bf 659}, 74 (2008).

\bibitem{prd76_122002}
M. Ablikim {\it et al.} (BES Collaboration), Phys. Rev. D
{\bf 76}, 122002 (2007).

\bibitem{prl97_121801}
M. Ablikim {\it et al.} (BES Collaboration), Phys. Rev. Lett.
{\bf 97}, 121801 (2006).

\bibitem{plb641_145}
M. Ablikim {\it et al.} (BES Collaboration), Phys. Lett. B
{\bf 641}, 145 (2006).

\bibitem{prl104_159901}
D. Besson {\it et al.} (CLEO Collaboration), Phys. Rev. Lett.
{\bf 104}, 159901 (2010).

\bibitem{prl96_092002}
D. Besson {\it et al.} (CLEO Collaboration), Phys. Rev. Lett.
{\bf 96}, 092002 (2006).

\bibitem{plb605_63}
J. Z. Bai {\it et al.} (BES Collaboration), Phys. Lett. B
{\bf 605}, 63 (2005).

\bibitem{prl96_082004} 
N. E. Adam {\it et al.} (CLEO Collaboration), Phys. Rev. Lett.
{\bf 96}, 082004 (2006).

\bibitem{prl96_182002} 
T. E. Coan {\it et al.} (CLEO Collaboration), Phys. Rev. Lett.
{\bf 96}, 182002 (2006).

\bibitem{prd74_031106} 
R. A. Briere {\it et al.} (CLEO Collaboration), Phys. Rev. D
{\bf 74}, 031106 (2006).

\bibitem{prd73_012002}
G. S. Adams {\it et al.} (CLEO Collaboration), Phys. Rev. D
{\bf 73}, 012002 (2006).

\bibitem{pdg2012}
J. Beringer {\it et al.} (Particle Data Group), Phys. Rev. D
{\bf 86}, 010001 (2012).

\bibitem{prd86_071101}
M. Ablikim {\it et al.}, (BESIII Collaboration), Phys. Rev. D
{\bf 86}, 071101 (2012).

\bibitem{prd87_112011}
M. Ablikim {\it et al.} (BESIII Collaboration), Phys. Rev. D
{\bf 87}, 112011 (2013).

\bibitem{prl91_022001}
J.Z. Bai {\it et al.} (BES Collaboration), Phys. Rev. Lett.
{\bf 91}, 022001 (2003).

\bibitem{cpc34_421}
M. Ablikim {\it et al.} (BESIII Collaboration), Chinese Phys. C
{\bf 34}, 421 (2010).

\bibitem{prd82_092002}
J.P. Alexander {\it et al.} (CLEO Collaboration), Phys. Rev. D
{\bf 82}, 092002 (2010).

\bibitem{prl95_262001}
M. Ablikim {\it et al.} (BES Collaboration), Phys. Rev. Lett.
{\bf 95}, 262001 (2005).

\bibitem{prl106_072002}
M. Ablikim {\it et al.} (BESIII Collaboration), Phys. Rev. Lett.
{\bf 106}, 072002 (2011).

\bibitem{prl108_112003}
M. Ablikim {\it et al.} (BESIII Collaboration), Phys. Rev. Lett.
{\bf 108}, 112003 (2012).

\bibitem{prd73_014516}
Y. Chen {\it et al.}, Phys. Rev. D {\bf 73}, 014516 (2006).

\bibitem{jhep_1210_170}
E. Gregory {\it et al.}, JHEP {\bf 73}, 170 (2012).

\bibitem{prd87_092009}
M. Ablikim {\it et al.}, (BESIII Collaboration), Phys. Rev. D
{\bf 87}, 092009 (2013).

\end{thebibliography}
\end{document}